# Ambient Electro-Synthesis of Ammonia - Electrode Porosity and Composition Engineering

Hong Wang‡*, Lu Wang‡, Qiang Wang, Shuyang Ye, Wei Sun, Yue Shao, Zhiping Jiang, Qiao Qiao, Yimei Zhu, Pengfei Song, Debao Li, Le He, Xiaohong Zhang, Jiayin Yuan*, Tom Wu, Geoffrey A. Ozin*

**Abstract:** Ammonia, key precursor for fertilizer production, convenient hydrogen carrier and emerging clean fuel, plays a pivotal role in sustaining life on earth. Currently, the main route for $NH_3$ synthesis is via the heterogeneous catalytic Haber-Bosch process ($N_2+3H_2 \rightarrow 2NH_3$), which proceeds under extreme conditions of temperature and pressure with a very large carbon footprint. Herein we report that a pristine nitrogen-doped nanoporous graphitic carbon membrane (NCM) can electrochemically convert $N_2$ into $NH_3$ in an aqueous acidic solution under ambient conditions. The Faradaic efficiency and rate of production of $NH_3$ on the NCM electrode reach 5.2% and 0.08 g $m^{-2}$ $h^{-1}$, respectively. After functionalization of the NCM with Au nanoparticles (Au NPs) these performance metrics are dramatically enhanced to 22% and 0.36 g $m^{-2}$ $h^{-1}$, respectively. These efficiencies and rates for the production of $NH_3$ at room temperature and atmospheric pressure are unprecedented. As this system offers the potential to be scaled to industrial proportions there is a high likelihood it might displace the century old Haber-Bosch process.

Fixation of $N_2$ to $NH_3$ is an essential process for maintaining life on earth[1-4]. Currently, $NH_3$ production is dominated by the Haber–Bosch process. It operates under conditions of high temperature 400–500 °C and pressure 200–250 bar, and its production has a huge carbon footprint[5]. The $H_2$ precursor, usually obtained by steam reforming of methane, also has a very large carbon footprint. Notably, the entire energy required to prepare the reagents and to operate the Haber-Bosch process amounts to 1–3% of the global energy supply[6]. In stark contrast, in the natural world, plants and bacteria have been producing $NH_3$ from $N_2$ and solvated protons under ambient conditions, enabled by the FeMo cofactor of the metalloenzyme nitrogenase ($N_2 + 6H^+ 6e^- \rightarrow 2NH_3$)[7-8]. Inspired by this biological nitrogen fixation process, intensive efforts have been devoted to finding ways to mimic the process under similarly mild conditions.

To this end, the electrocatalytic $N_2$ reduction reaction (NRR) conducted in an aqueous media has recently been receiving increasing attention. This approach offers multiple merits: (i) use of water as the hydrogen source, (ii) operation under ambient conditions, and (iii) utilization of renewable electricity to drive the process[9-10]. Nonetheless, the extremely high bond energy of the $N_2$ molecule, 940.95 kJ $mol^{-1}$ together with its lack of a permanent dipole, makes running the NRR under mild conditions extremely challenging[11]. The quest for suitable electrocatalysts and electrolytes for the NRR represents one of the most active areas of materials and energy research[12-17]. Among the electrocatalysts and electrolytes for the NRR, the most efficient ones rely on high temperature reactions (T> 200 °C) to favor the reaction thermodynamics. For example, Licht et al. found that an electrochemical cell constructed with a Ni electrode and a ternary molten hydroxide (KOH/NaOH/CsOH) suspension of nano-$Fe_2O_3$ could produce $NH_3$ at a coulombic efficiency of 35% by electrolysis of air and steam at 200 °C[18]. Marnellos et al. demonstrated that a Pd electrode in combination with $ScCe_{0.95}Yb_{0.05}O_{3-\alpha}$ as a solid-state proton conductor could generate $NH_3$ with a Faradaic efficiency of 78% by electrolysis of $N_2$ and $H_2$ at 570 °C[19].

Herein we report for the first time that hierarchically structured nitrogen-doped nanoporous carbon membranes (NCMs) can electrochemically convert $N_2$ into $NH_3$ at room temperature and atmospheric pressure in an acidic aqueous solution. The Faradaic efficiency and rate of $NH_3$ production using the metal-free NCM electrode in 0.1 M HCl solution are as high as 5.2% and 0.08 g $m^{-2}$ $h^{-1}$, respectively. Upon functionalization of the NCM electrode with Au nanoparticles (Au NPs), the efficiency and rate are boosted to a remarkable 22% and 0.36 g $m^{-2}$ $h^{-1}$, respectively. These performance metrics are unprecedented for the electrocatalytic production of $NH_3$ from $N_2$ under ambient conditions.

To amplify, **Fig. 1a-c** shows the synthetic protocol for making NCMs (see details in Supplementary information). To begin, a homogeneous dispersion of multi-wall CNTs was prepared by sonicating CNTs in a solution of poly(acrylic acid) (PAA) and a poly(ionic liquid) named poly[1-cyanomethyl-3-vinylimidazolium bis(trifluoromethanesulfonyl)imide] (PCMVImTf$_2$N) in N,N-dimethyl formamide (**Fig. 1a**). The chemical structures of PCMVImTf$_2$N and PAA are shown in **Supplementary Fig. 1**. The stable polymer/CNT dispersion was cast onto a glass plate, dried at 80 °C into a sticky black film, and finally immersed in an aqueous $NH_3$ solution (0.1wt %) to build up a nanoporous polymer/CNT hybrid membrane (**Fig. 1b, Supplementary Fig. 2**). Afterwards, pyrolysis treatment of the hybrid membrane at 900 °C under $N_2$ leads to the targeted NCM (**Fig. 1c**). The porous surface can be seen from the top view of the NCM in a SEM image (**Fig. 1d**). **Fig. 1e** shows that a three-dimensionally interconnected macroporous architecture was created along the cross-section. A high-magnification SEM image (**Fig. 1f**) reveals that CNTs are uniformly embedded in the NCM, which is expected owing to their uniform dispersion in the PCMVImTf$_2$N/PAA DMF solution.

[*] Dr. H. Wang, Mr. Y. Shao, Miss. Z. Jiang
Institute of Polymer Chemistry, College of Chemistry, Nankai University, Tianjin, 300071, P. R. China
E-mail: hongwang1104@nankai.edu.cn
Dr. L. Wang, Mr. S. Ye, Dr. W. Sun, Prof. G. A. Ozin
E-mail: gozin@chem.utoronto.ca
Dr. L. Wang, Prof. L. He, Prof. X. Zhang
Institute of Functional Nano & Soft Materials, Soochow University, Suzhou, Jiangsu 215123, P. R. China
Dr. Q. Wang, Dr. D. Li
State Key Laboratory of Coal Conversion, Institute of Coal Chemistry, The Chinese Academy of Sciences, Taiyuan 030001, China
Dr. Qiao, Dr. Y. Zhu
Department of Physics, Temple University, Philadelphia, Pennsylvania 19122, USA
Prof. P. Song
College of Chemistry and Chemical Engineering, Northwest Normal University, Lanzhou 730070, P.R. China
Prof. J. Yuan
Department of Materials and Environmental Chemistry, Stockholm University, 10691, Stockholm, Sweden
Email: jiayin.yuan@mmk.su.se
Prof. T. Wu
School of Materials Science and Engineering, UNSW Australia, Kensington Campus Building E10, Sydney, NSW 2052, Australia
‡Authors contributing equally to the work.

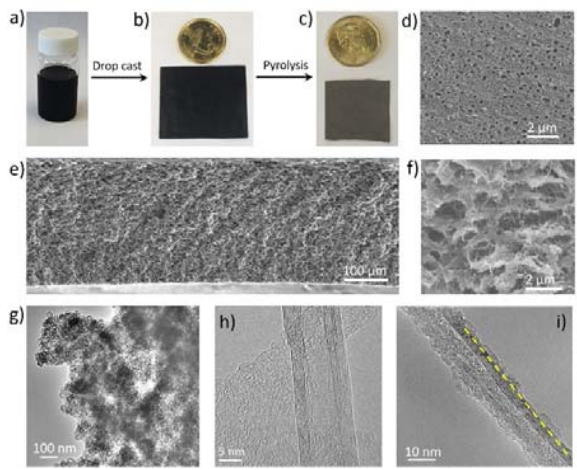

**Figure 1.** a)-c), Scheme illustrating the synthetic route to the NCM. Image (a) depicts a homogeneous dispersion of CNTs in a solution of PCMVImTf$_2$N and PAA in DMF; Image (b) is a PCMVImTf$_2$N/PAA/CNT hybrid membrane; Image (c) is a NCM prepared from pyrolysis of sample in (b); d), SEM image of the top surface of the NCM. e), f), Low- and high-magnification cross-section SEM images of the NCM, respectively. g, a HRTEM image of NCM. h), a single native CNT. i), the core-shell structure of a CNT-NC heterojunction. The yellow arrow directed area represents N-doped carbon sheath attached to a CNT.

The microstructure of the NCM was further analyzed by high-resolution transmission electron microscopy (HRTEM), in which CNTs embedded in the nanoporous carbon membrane matrix are clearly identified (**Fig. 1g, Supplementary Fig. 3**). The well-defined lattice spacing of 0.34 nm indicates that the NCM membrane contains highly organized graphitic domains (**Supplementary Fig. 4**). The individual graphitic layers can be seen to bend due to a doping effect of nitrogen atoms, and extend practically across the entire membrane. The content of nitrogen is 9.0 wt. % as determined by elemental analysis. **Fig. 1h** shows a HRTEM image of a native CNT, which is typically composed of 7-12 layers with an outer diameter of 5-10 nm. Notably, in **Fig. 1i** and **Supplementary Fig. 5**, a thin rough sheath is formed on the CNT wall (see the area indicated by the yellow arrow), referred to as a CNT-NC core-shell microstructure. We previously demonstrated that the imidazolium cations, a major component in PCMVImTf$_2$N polymer, could attach to the graphitic CNT surface owing to the well-known cation-π interaction for dispersion purposes[20]. It has recently been reported that the CNT surface could template and catalyze the pyrolysis of ionic liquid species to form structural motifs different from their bulk carbonization[21]. The newly formed sheath of N-doped carbon on the surface of CNT is rich in defects (dangling bonds) and short-range ordered micropores. Specifically, the heteroatoms, through templating interactions, favorably align themselves at the CNT surface for exposure to reagents, thereby enabling high electron transfer efficiency via charge transfer interactions, which in turn will modify catalytic activity.

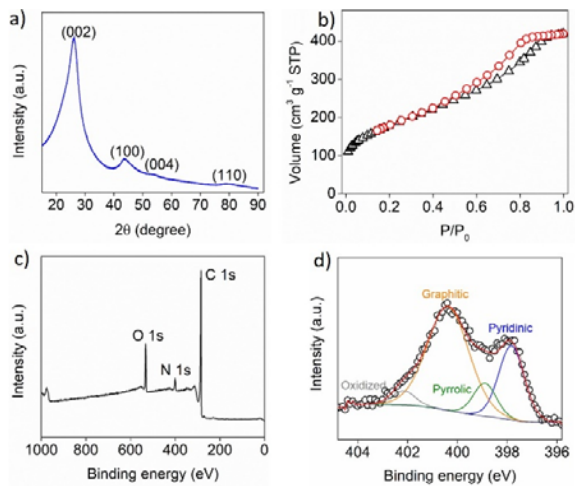

**Figure 2.** a), XRD pattern of the NCM. b), BET specific surface area of the NCM. c), XPS spectrum of the NCM. d), High-resolution N1 XPS spectrum of the NCM and its deconvolution into four chemically distinct N forms.

**Fig. 2a** shows the X-ray diffraction (XRD) pattern of the NCM. Sharp diffraction peaks at 26°, 44°, 53° and 80° are observed and attributed to the (002), (100), (004) and (112) reflections of a graphitic carbon, respectively. Such a graphitic structure of the NCM endows it with high electronic conductivity (134 S cm$^{-1}$ at 298 K)[22], which favors fast charge transport, a mandatory requirement for efficient electrocatalysis.

Specific surface area plays an essential role in optimizing catalytic activity of heterogeneous catalysts. The Brunauer-Emmett-Teller (BET) specific surface area of NCM (**Fig. 2b**) is measured to be 432 m$^2$/g with a pore volume of 0.58 cm$^3$/g. It is clear from the imaging and pore size distribution (**Supplementary Fig. 6**) studies that the pore architecture of the NCM membrane is hierarchical in nature, comprising of macropores seen in the cross-sectional image with pores traversing the entire micro- to meso- to macropore range. In such hierarchically porous membrane architectures, micropores and small mesopores are beneficial to provide large and accessible surface area. Plus the large mesopores and macropores form interconnected three-dimensional networks, which can serve as transport highways to accelerate mass diffusion and promote electron exchange efficiency and catalytic activity. Furthermore, the hierarchical porous architecture and favorable electrical conductivity of the NCM allow it to function as a diffusion electrode to enhance the three-phase contact and charge transport between the heterogeneous electrocatalyst, aqueous electrolyte and gaseous reactants to optimize the electrocatalytic reaction rate.

X-ray photoelectron spectroscopy (XPS) show diagnostic C, N and O peaks in the NCM (**Fig. 2c**), and provide evidence that the NCM is metal-free. An analysis of different N species in the NCM structure is presented in **Fig. 2d**. The N 1s XPS spectrum shows that N in the carbon framework exists mainly in the pyridinic (398.0 eV), pyrrolic (398.6 eV), graphitic (400.2 eV) and oxidized (402.2 eV) forms with corresponding abundances of 32.2%, 12.9%, 48.1%, 6.8%, respectively. It is relevant that the high pyridinic and pyrrolic N content in NCM, which we attribute to surface templating of the condensation reactions, co-exists with coupled edge termination of graphitic layers by pyridinic-pyrrolic units. It is considered that pyridinic and pyrrolic N atoms are the catalytic active sites in N-doped carbons[23]. Previous reports have demonstrated that dangling bonds are key to NRR activity[13]. Considering that abundant pyridinic and pyrrolic N atoms, considered as dangling bonds, exist in the NCM, in what follows, we will discuss the NRR performance of the NCM in an acidic solution under ambient conditions.

The NCM was directly utilized as the working electrode for the NRR in 0.1 M HCl (pH =1) electrolyte at ambient conditions. Noticeably, a polymer binder, which is mandatory for use of powdered electrocatalysts is not needed to construct the working electrode due to the structural connectivity of the NCM. The $N_2$ is supplied in a feed gas stream to the diffusion-type porous NCM cathode, where protons ($H^+$) are transported through the electrolyte to react with $N_2$ to produce $NH_3$ ($N_2 + 6H^+ + 6e^- \rightarrow 2NH_3$). We firstly investigated the overpotential required to achieve the maximum Faradaic efficiency of $NH_3$ synthesis for the NCM electrode from 0 V to -0.4 V (vs RHE). The system was tested for an extended period of 3 hours (**Supplementary Fig. 7**). The yield of $NH_3$ produced by the NRR on the NCM electrode was measured by using the indophenol blue method[3,24] (**Supplementary Fig. 8-9**). The achievable maximum Faradaic efficiency of 5.2% for $NH_3$ was reached at -0.2 V (vs RHE) (**Fig. 3a**) and the highest rate of $NH_3$ is 0.08 g m$^{-2}$ h$^{-1}$ at -0.3 V (vs RHE) (**Fig. 4b**, **Supplementary Fig. 10**). As shown in **Fig. 3b**, the rate of $NH_3$ decreases significantly beyond -0.3 V (vs RHE), which is attributed to a competitive reduction of $N_2$ and hydrogen species on the electrode surface. In the case of the Haber-Bosch $NH_3$ synthesis, $N_2H_4$ is the major by-product. Notably, in our NCM based electrochemical process, $N_2H_4$ was not detected (**Supplementary Fig. 11**), indicative of a 100% selectivity of the NCM for reduction of $N_2$ to $NH_3$.

In control experiments, when $N_2$ was replaced by argon, while keeping other reaction parameters unchanged, $NH_3$ could not be identified in the electrolyte. The same was observed when the NCM was replaced by a carbon paper or pristine CNTs (**Supplementary Fig. 12-14**). These control experiments confirmed that $NH_3$ was produced exclusively from the NRR reaction catalysed by the NCM electrode. In addition, XPS spectra of the NCM after a 4-day stability test were recorded (**Supplementary Fig. 15**), which show no difference compared to the pristine NCM. Such robust electrochemical stability is likely attributed to the nitrogen doping effect, which improves the electrochemical stability and resistance against oxidation by modifying the electronic band structure of the graphitic carbons.

Note that the NCM was fabricated by direct pyrolysis of a polyelectrolyte complex membrane. It is well known that polyelectrolytes are capable of binding and immobilizing metal ions, salts and nanoparticles[25]. This knowledge inspired us to functionalize the NCM with metal nanoparticles via doping the polymer membrane with metal species before carbonization. In this context we co-assembled Au nanoparticles (Au NPs) with the NCM to amplify the performance of the NCM for the NRR. The detailed synthetic procedure and structural characterizations, including digital photograph, XRD, SEM, TEM, Au NPs size distribution, HRTEM and elemental mappings of the NCM-metal nanoparticle hybrid (NCM-Au NPs) are provided in the Supplementary information and **Supplementary Fig.16-21**.

The loading of Au within the NCM was 6.03 wt % as detected by inductively coupled plasma−atomic emission spectra (ICP−AES). The composite NCM-Au NPs electrode for NRR testing was the same as the pristine NCM (**Supplementary Fig. 22**). The optimised Faradaic efficiency and rate of NCM-Au NPs in conversion of $N_2$ to $NH_3$ was as high as 22% at -0.1 V vs RHE (**Fig. 3c**) and 0.36 g m$^{-2}$ h$^{-1}$ at -0.2 V vs RHE (**Fig. 3d**, **Supplementary Fig. 23**) respectively. Note that there is no $NH_4Cl$ produced in electrolyte without employing applied potential (**Supplementary Fig. 24**). These are the highest values ever reported for $NH_3$ production at ambient conditions. Similar to the NCM electrode, there is no $N_2H_4$ found using the NCM-Au NPs electrode (**Supplementary Fig. 25**), certifying the 100% selectivity of NCM-Au NPs in this process. Furthermore, to ensure the $NH_3$ product of the reactions did not originate from adventitious N residues in our samples, isotopically labelled $^{15}N_2$ authenticated the origin of the products of the NRR. We bubbled 20 mL of 0.1 M HCl electrolyte with $N_2$ using a large piece of NCM-Au NPs (6.5 x 3 cm$^2$) as the working electrode, Ag/AgCl as reference electrode and Pt as counter electrode under -0.2 V vs RHE (**Fig. 4a**), after reacting for 28 h, the white and slightly yellow product was collected by freeze-drying of the electrolyte. The 1H-NMR spectrum of produced $^{15}NH_4Cl$ shows the doublet peaks (6.89 ppm, $J^1_{NH}$ = 72.2 Hz) (**Fig. 4b**), consistent with previously reported[2], which clearly demonstrated that NCM-Au can efficiently transfer $N_2$ into $NH_3$, being potential to be scaled to industrial proportions there is a high likelihood it might displace the century old Haber-Bosch process.

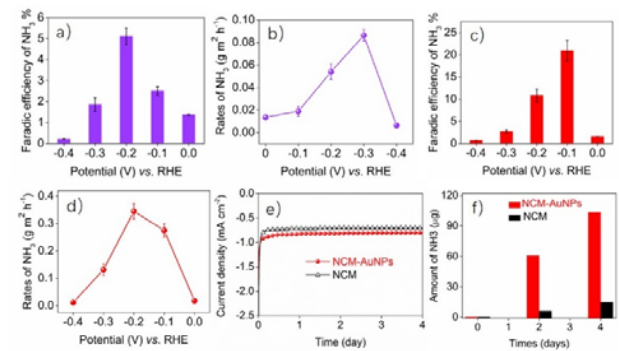

**Figure 3.** a), Faradaic efficiencies for $NH_3$ production vs applied potential at the NCM electrode (vs RHE). b), the rates of $NH_3$ production on the NCM electrode at applied potentials (vs RHE). c), Faradaic efficiencies for $NH_3$ production vs the applied potential at the NCM-Au NPs electrode (vs RHE). d), the rates of $NH_3$ production on the NCM-Au NPs electrode at applied potentials (vs RHE). e), Chrono-amperometry results at the corresponding potentials (in a and c) with the highest Faradaic efficiencies. f), the yields of $NH_3$ production at regular interval times on the NCM and NCM-Au NPs electrodes during a long-term operational stability test.

To gain additional insight into the catalytic kinetics of the NRR, electrochemical impedance spectroscopy measurements were performed at various overpotentials. As shown in Fig. SXX, Nyquist plots were used to determine the series resistance ($R_s$), pore resistance ($R_p$) and charge-transfer resistance ($R_{ct}$). The overpotential-independent behaviours of $R_p$ for NRR (~0.4 Ω/cm$^2$) indicate the robust nature of the hierarchical pore structures of the catalyst even at high overpotentials, and these structures also serve as efficient channels for mass transport to access the exposed active sites. In addition, NCM-Au NPs exhibited a much lower $R_{ct}$ for NRR (0.047 $\Omega/cm^2$ at 227 mV), indicating fast charge transfer during the reactions due to the highly conductive nature of the NCM-Au NPs **Supplementary Fig. 26**.

The stability of an electrocatalyst is key for its practical applications. Both NCM and NCM-Au NPs membranes were operated for 4 days, as shown in **Fig. 3e**. No decay of activity was observed, indicating excellent electrochemical durability for both systems. The yield of $NH_3$ produced on both NCM and NCM-Au NPs electrodes were measured at regular intervals during the stability test (**Supplementary Fig. 27-28**), and as shown in **Fig. 3f**, the yield of $NH_3$ actually was observed to increase with reaction time. This impressive performance clearly demonstrates the robustness of NCM and NCM-Au NPs electrodes for NRR.

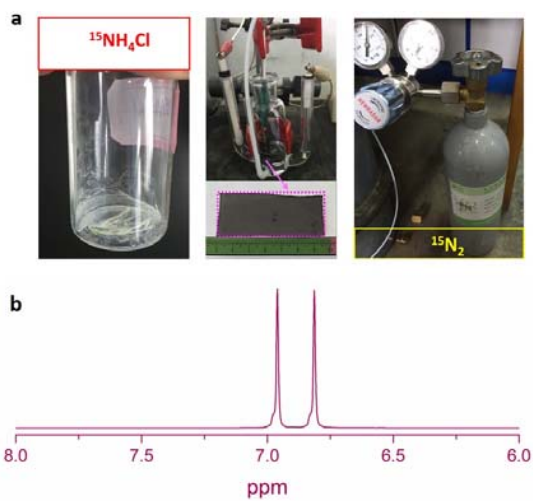

**Figure 4.** a), Photograph of equipment used for producing $^{15}NH_4Cl$. b), The $^1H$-NMR spectrum of isolated $^{15}NH_4Cl$ in the above setup.

Although previous theoretical predictions and experimental tests showed that Au nanomaterials are NRR-active, their $N_2$ reduction yield and product selectivity are limited[26]. In stark contrast, our NCM-Au NPs electrode exhibits unprecedented NRR performance. Firstly, Au NPs homogeneously dispersed and highly embedded within the NCM, could improve carbon–metal interactions and increase the number of chemically active sites[27]. While still under debate, it is generally agreed that the mechanism of the heterogeneously catalysed NRR can be either associative or dissociative[7,28] (**Supplementary Fig. 29**). For both mechanisms, the first step is $N_2$ binding to the surface of the heterogeneous catalyst. In our case, the heterojunction between the Au NPs and semiconductor-like NCM, verified by temperature-dependent conductivity measurements, as shown in **Supplementary Fig. 30**, creates a rectifying effect[29]. At the interface of Au/NCM, electrons are expected to flow from NCM to Au (**Supplementary Fig. 31**), considering the $E_f$ of Au and NCM are -5.0 eV and -4.5 eV, respectively. In this case, the positively charged NCM surface so formed may adsorb $N_2$ more strongly and thereby improve the NRR activity[28]. Together, the synergistic charge-transfer between Au NPs and NCM creates excellent NRR performance of the NCM-Au NPs electrode (Table S1).

Considering the excellent conversion efficiency and selectivity of $N_2$ to $NH_3$ by NCM and NCM-AuNP electrodes, and the straightforward method for production of the NCM membranes at scale, the advance described herein provides exciting opportunities for developing a highly efficient, industrial electrochemical process for producing $NH_3$ from abundant $N_2$ and $H_2O$ under ambient temperature and pressure conditions. With further optimization of this electrocatalytic system there is an excellent chance it will be able to outcompete the century old heterogeneous catalytic Haber-Bosch process that operates under extreme conditions of temperature and pressure.

## Acknowledgements

G.A.O. is a Government of Canada Research Chair in Materials Chemistry and Nanochemistry. Financial support for this work was provided by the Ontario Ministry of Research Innovation (MRI); Ministry of Economic Development, Employment and Infrastructure (MEDI); Ministry of the Environment and Climate Change; Connaught Innovation Fund; Connaught Global Challenge Fund; and the Natural Sciences and Engineering Research Council of Canada (NSERC). J. Y. is grateful for financial support from the ERC Starting Grant NAPOLI - 639720. H. W. acknowledges the financial support from the Nankai University.

**Keywords:** Poly(ionic liquid) • Porous carbon membrane • Functionalization • Electrocatalysis • Nitrogen fixation •

[1]   R. Lan, J. T. S. Irvine, S. Tao, *Sci. Rep.* **2013**, *3*, 1145.
[2]   C. Liu, K. K. Sakimoto, B. C. Colón, P. A. Silver, D. G. Nocera, *Proc. Natl. Acad. Sci.* **2017**, 114, 6450–6455.
[3]   D. Zhu, L. Zhang, R. E. Ruther, Hamers, R. J. *Nat. Mater.* **2013**, *12*, 837–841.
[4]   J. H. Montoya, C. Tsai, A. Vojvodic, J. K. Norskov, *ChemSusChem.* **2015**, *8*, 2180–2186.
[5]   D. Tilman, K. G. Cassman, P. A. Matson, R. Naylor, S. Polasky, *Nature* **2002**, *418*, 671–677.
[6]   M. A. H. J. V. Kessel, D. R. Speth, M. Albertsen, P. H. Nielsen, H. J. M. Op den Camp, B. Kartal, M. S. M. Jetten, S. Lücker, *Nature*, **2015**, *528*, 555-559.
[7]   C. D. Van der Ham, M. T. M. Koper, D. G. H. Hetterscheid, *Chem. Soc. Rev.* **2014**, *43*, 5183–5191.
[8]   K. A. Brown, D. F. Harris, M. B. Wilker, A. Rasmussen, N. Khadka, H. Hamby, S. Keable, G. Dukovic, J. W. Peters, L. C. Seefeldt, P. W. King, *Science* **2016**, *352*, 448-450.
[9]   S. Gao, Y. Lin, X. Jiao, Y. Sun, Q. Luo, W. Zhang, D. Li, J. Yang, Y. Xie, *Nature*, **2016**, *529*, 68-71.
[10]  C. Costentin, M. Robert, J.-M. Saveant, *Chem. Soc. Rev.* **2013**, *42*, 2423–2436.
[11]  K. Honkala, A. Hellman, I. N. Remediakis, A. Logadottir, A. Carlsson, S. Dahl, C. H. Christensen, J. K. Nørskov,  Science **2005**, *307*, 555–558.
[12]  N. Furuya, H. J. Yoshiba, *Electroanal. Chem.* **1990**, *291*, 269–272.
[13]  S.-J. Li, D. Bao, M.-M. Shi, B.-R. Wulan, J.-M. Yan, Q. Jiang, *Adv. Mater.* **2017**, *33*, 1700001.
[14]   M. A. Shipman, M. D. Symes, *Catal. Today* **2017**, *286*, 57–68.
[15]  V. Kyriakou, I. Garagounis, E. Vasileiou, A. Vourros, M. Stoukides, *Catal. Today* **2017**, *286*, 2–13.
[16]  G.-F. Chen, X.Cao, S. Wu, X. Zeng, L.-X. Ding, M. Zhu, H. Wang, J. *Am. Chem. Soc.* **2017**, *139*, 9771–9774.
[17]  J. Zhao, Z. Chen, *J. Am. Chem. Soc.* **2017**, *139*, 12480–12487.
[18]   S. Licht, B. Cui, B. Wang, F.-F. Li, J. Lau, S. Liu, *Science* **2014**, *345*, 637–640.
[19]  G. Marnellos, M.  Stoukides, *Science* **1998**, *282*, 98–100.

[20]   S. Soll, M. Antonietti, J. Yuan, *ACS Macro Lett.* **2012**, *1*, 84–87.

[21]   Y. Ding, X. Sun, L. Zhang, S. Mao, Z. Xie, Z.-W. Liu, D. S. Su, *Angew. Chem. Int. Ed.* **2015**, *54*, 231 –235

[22]   H. Wang, J. Jia, Q. Wang, D. Li, S. Min, C. Qiana, L. Wang, C. Ma, T. Wu, J. Yuan, M. Antonietti, G. A. Ozin, *Angew. Chem. Int. Ed.* **2017**, *56*, 7847–7852.

[23]   W. Ouyang, D. Zeng, X. Yu, F. Xie, W. Zhang, J. Chen, J. Yan, F. Xie, L. Wang, H. Meng, D. Yuan, *Int. J. Hydrogen. Energy.* **2014**, *39*, 15996.

[24]   D. F. Bolt, Colorimetric Determination of Nonmetals, Wiley, 1978, Vol. 2.

[25]   M. Schrinner, M. Ballauff, Y. Talmon, Y. Kauffmann, J. Thun, M. Möller, J. Breu, *Science* **2009**, 323, 617–620.

[26]   D. Bao, Q. Zhang, F.-L. Meng, H.-X. Zhong, M.-M. Shi, Y. Zhang, J-M. Yan, Q. Jiang, X.-B. Zhang, *Adv. Mater.* **2017**, 29, 1604799.

[27]   R. J. White, R. Luque, V. L. Budarin, J. H. Clark, D. J. Macquarrie, *Chem. Soc. Rev.* **2009**, *38*, 481–494.

[28]   C. Guo, J. Ran, A. Vasileffa, S.-Z. Qiao, *Energy Environ. Sci.* **2018**,*11*, 45-56.

[29]   X.-H. Li, M. Antonietti, *Chem. Soc. Rev.* **2013**, *42*, 6593–6604.

Supplementary Information for

## Efficient Electrocatalytic Reduction of CO2 by Nitrogen-Doped Nanoporous Carbon/Carbon Nanotube Membranes: A Step Towards the Electrochemical CO2 Refinery

Hong Wang[1][‡][*], Lu Wang[‡][2,3], Qiang Wang[4], Shuyang Ye[2], Wei Sun[2], Yue Shao[1], Zhiping Jiang[1], Qiao Qiao[5,6], Yimei Zhu[6], Pengfei Song[7], Debao Li[4], Le He[3], Xiaohong Zhang[3], Jiayin Yuan[*][8], Tom Wu[9], Geoffrey A. Ozin[2][*]

[1] Key Laboratory of Functional Polymer Materials, Ministry of Education, Institute of Polymer Chemistry, Nankai University, Tianjin, 300071, P. R. China;

[2] Materials Chemistry and Nanochemistry Research Group, Solar Fuels Cluster, Departments of Chemistry, University of Toronto, 80 St. George Street, Toronto, Ontario M5S3H6, Canada;

[3]Institute of Functional Nano & Soft Materials (FUNSOM), Jiangsu Key Laboratory for Carbon-Based Functional Materials & Devices, Soochow University, 199 Ren'ai Road, Suzhou, Jiangsu 215123, People's Republic of China;

[4] State Key Laboratory of Coal Conversion, Institute of Coal Chemistry, The Chinese Academy of Sciences, Taiyuan 030001, China;

[5]Department of Physics, Temple University, Philadelphia, Pennsylvania 19122, USA;

[6]Condensed Matter Physics and Materials Science Department, Brookhaven National Laboratory, Upton, New York 11973, USA;

[7]College of Chemistry and Chemical Engineering, Northwest Normal University, Lanzhou 730070, China; [8]Department of Materials and Environmental Chemistry, Stockholm University, 10691, Stockholm, Sweden;

[9] School of Materials Science and Engineering, UNSW Australia, Kensington Campus Building E10, Rm345, Sydney, NSW 2052, Australia;

Corresponding author
Email: hongwang0104@nankai.edu.cn;   jiayin.yuan@mmk.su.se;n   gozin@chem.utoronto.ca;

## 1. Materials

1-Vinylimidazole (Aldrich 99%), 2,2'-azobis(2-methylpropionitrile) (AIBN, 98%), bromoacetonitrile (Aldrich 97%), chloroauric acid (HAuCl₄, Aldrich 99.99%), Sodium hydroxide (NaOH Aldrich 98%), salicylic acid (Aldrich 99%), sodium citrate (99.5%), Sodium hypochlorite (NaClO, Aldrich 99%), sodium nitroferricyanide (Na[Fe(NO)(CN)₅], Aldrich 99%), para-(dimethylamino) benzaldehyde ((CH₃)₂NC₆H₄CHO, Aldrich 99%) and bis(trifluoromethane sulfonyl)imide lithium salt (Aldrich 99%) were used as received without further purifications. Dimethyl sulfoxide (DMSO), dimethyl formamide (DMF), methanol (MeOH), Ethanol (EtOH), tetrahydrofuran (THF) were of analytic grade. Potassium bicarbonate (KHCO₃) was of analytic grade and purchased from Sigma Aldrich. Poly(acrylic acid) (PAA) MW: 130,000 g/mol, was obtained from Sigma Aldrich. $^{15}N_2$ (WUHAN NEWRADAR SPECIAL GAS CO., LTD., Purity 99.13%) was used as received. Multi-walled carbon nanotube (≥ 98 %, O.D x I.D x L.D 10 nm ± 1 nm x 4.5 nm ± 0.5 nm x 3 ~6 μm) was purchase from Sigma Aldrich and purified by the following procedure: 10 g of MWCNT were mixed with 500 ml of acid (68%), stirred and heated up under reflux at around 100 °C for 20 h. After the reaction the gaseous supernatant was purged with nitrogen to remove acidic vapor for better handling. The material was filtered and washed extensively in a washing cell over night with distilled water to remove residual acid and impurities like iron. After that, the black wet CNT was dried in drying oven at 100 ºC for 12 h. The as-obtained CNT powder then was annealed in furnace at 1000 ºC for 2 h (N₂ condition). Poly[1-cyanomethyl-3-vinylimidazolium bis(trifluoromethanesulfonyl) imide] (abbreviation: PCMVImTf₂N) was synthesized according to reference (S*1*).

1.1 Preparation of the NCM

A homogeneous CNT dispersion was firstly accomplished by sonicating 0.3 g of purified CNTs in 3g of PCMVImTf$_2$N and 0.54g of PAA solution in 40 mL of DMF for 4 h. Owing to the strong polarization interactions between the imidazole units in PCMVImTf$_2$N and CNTs, PCMVImTf$_2$N specifically attaches to the surface of the CNTs. Then, centrifugation of the solution removes aggregations. The formed homogeneous dark solution was then cast onto a clean glass plate and dried at 80 °C for 3 h. Then the film was immersed in a 0.1 wt % aqueous NH$_3$ solution for 3 h. This process triggers charging of the PAA moieties and electrostatic crosslinking of the charged PAA with PCMVImTf$_2$N simultaneously, constructing the film on the nanoscale level, and formation of a stable porous structure within the film, which can be readily peeled off the substrate. Afterwards, pyrolysis of the film at 900 °C in pure N$_2$ leads to the hierarchically structured NCM.

1.2 Preparation of the NCM-Au NPs

Freshly prepared porous PCMVImTf$_2$N/PAA/CNTs membrane was placed in 200 mL of chloroauric acid aqueous solution (2 wt %) for 12 h. Afterward, PIL/PAA/CNTs-HAuCl$_4$ was taken out from the solution, washed with water, and dried at room temperature till constant weight. Finally, direct pyrolysis of PCMVImTf$_2$N /PAA/CNTs-HAuCl$_4$ membrane was carried out similarly to that of the NCM, leading to NCM-Au NPs membrane.

**2. Characterization**

X-ray diffraction was performed on a Bruker D2-Phaser X-ray diffractometer, using Cu Kα radiation at 30 kV. Nitrogen adsorption isotherms were obtained at 77 K using a Quantachrome Autosorb-1-C. The surface area of sample was determined using BET theory, and pore size distributions were determined with NLDFT. X-ray photoelectron spectroscopy (XPS) data were collected by an Axis Ultra instrument (Kratos Analytical) under ultrahigh vacuum ($<10^{-8}$ Torr) and by using a monochromatic Al Kα X-ray source. The adventitious carbon 1s peak was calibrated at 285 eV and used as an internal standard to compensate for any charging effects. All data analyses were carried out using the Multipak fitting program, and the binding energies were referenced to the NIST-XPS database and the Handbook of X-ray Photoelectron Spectroscopy. The amount of Au NPs in NCM-Au NPs was quantified by inductively coupled plasma-atomic emission spectroscopy (ICP−AES, Thermo Electron Corp. Adv. ER/S). Before the ICP measurements, the heterostructures were immersed in aqua regia with agitation for 12 h to dissolve the Au NPs. Transmission electron microscope (TEM) and high-resolution TEM (HRTEM) images, selected-area electron diffraction (SAED) patterns were taken on a Hitachi H-7000 transmission electron microscope at 100 kV. The $^1$H NMR spectrum was recorded on a Bruker 500 with Probe TXI at temperature of 25°C using 3 mm tube. The sample was dissolved in 1M HCl solution (D$_2$O/H$_2$O mixed solution).

**3. Electrochemical testing**

The electrochemical measurements were performed by electrochemical impedance spectroscopy (EIS) using a Biologic VMP3 potentiostat. The as-prepared NCMs or NCMs-Au NPs were used as the working electrode. A Pt and an Ag/AgCl (in saturated KCl solution) electrode were used as the counter and reference electrodes, respectively. The electrolyte is 0.1 M HCl solution with a pH value is 1. All the applied potentials were converted to reversible hydrogen electrode (RHE) potential scale (without the IR compensation) using E (vs. RHE) = E (vs. Ag/AgCl) + 0.197 V+0.0592 V x pH. A fixed volume of 20 mL electrolyte was used for electrochemical experiments (50 mL electrolyte was used for stability test). N$_2$ gas was continuously bubbled into the cathodic compartment at a rate of 10 mL min$^{-1}$. The electrode area was calculated from its surface area.

*Determination of ammonia*: Absolute calibration of the method was achieved using ammonium chloride solutions of known concentration as standards (S$2$). The production of ammonia was measured using the indophenol blue method. 2 ml aliquot of solution was removed from the reaction vessel. To this solution was added 2 ml of a 1M NaOH solution containing 5% salicylic acid and 5% sodium citrate (by weight), followed by addition of 1 ml of 0.05 M NaClO and 0.2 ml of an aqueous solution of 1% sodium nitroferricyanide (by weight). After 2 h, the absorption spectrum was measured using a Varian Cary 5000 spectrophotometer. The formation of indophenol blue was determined using the absorbance at a wavelength of 655 nm.

*Determination of hydrazine hydrate*: The hydrazine present in the electrolyte was estimated by Watt and Chrisp' method. (*S3*). A mixture of para-(dimethylamino) benzaldehyde (5.99 g), HCl (concentrated, 30 mL) and ethanol (300 mL) was used as a color reagent. Calibration curve was plotted as follow: First, preparing a series of reference solutions, by pipetting suitable volumes of the hydrazine hydrate-nitrogen 0.1 M HCl solution in colorimetric tubes; Second, making up to 5 mL with 0.1 M HCl solution; Third, adding 5 mL above prepared color reagent and stirring 10 min at room temperature. Fourth, the absorbance of the resulting solution was measured at 455 nm, and the yields of hydrazine were estimated from a standard curve using 5 mL residual electrolyte and 5 mL color reagent.

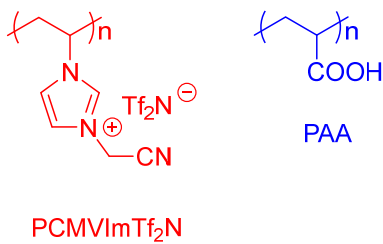

**Fig. S1. The chemical structure of PCMVImTf$_2$N left and PAA right.**

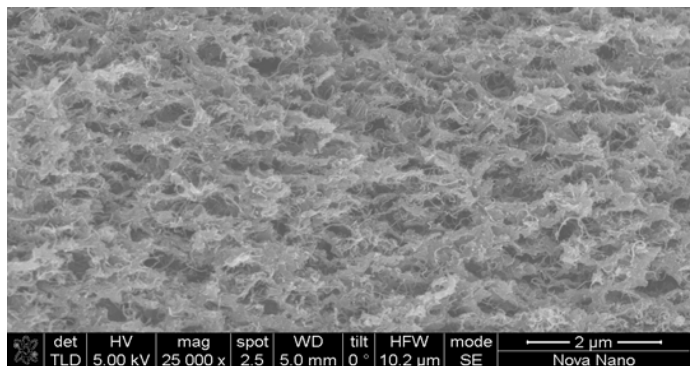

**Fig. S2. SEM image of the nanoporous PCMVImTf$_2$N/PAA /CNTs film membrane.**

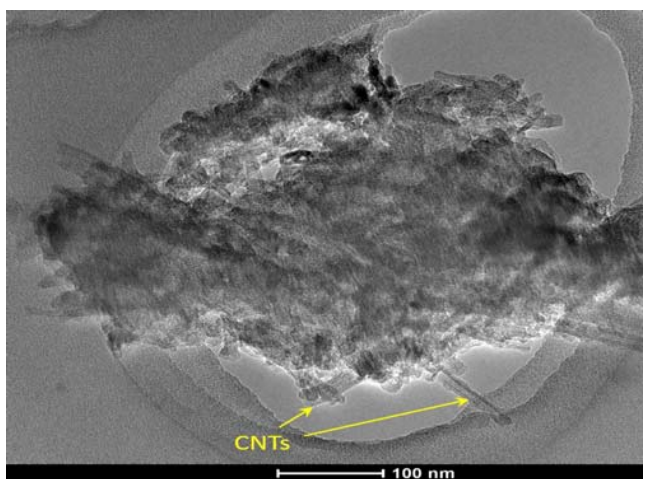

**Fig. S3. SEM image of the NCM.** CNTs uniformly embedded in a N-doped carbon membrane matrix are observable.

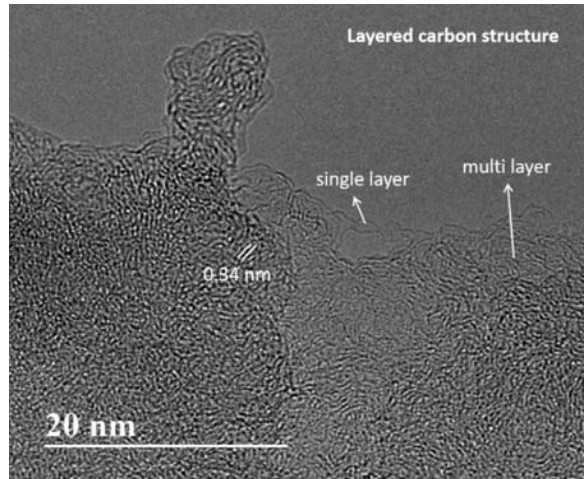

**Fig. S4. HRTEM image of the NCM**

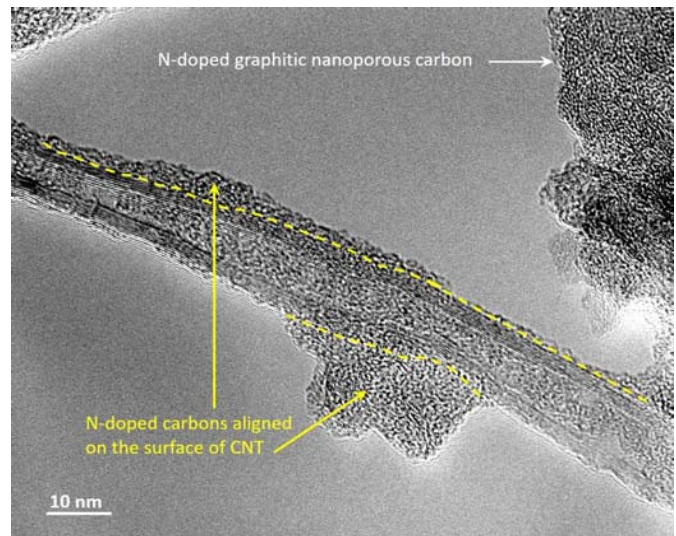

**Fig. S5. HRTEM image of the NCM.** The core-shell structure of a CNT-NC heterojunction can be clearly seen. The white arrow directed area represents N-doped carbon sheath attached to a CNT;

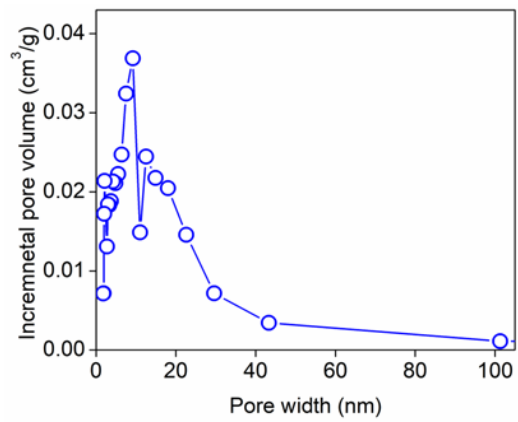

**Fig. S6. Pore size distribution curve of the NCM.**

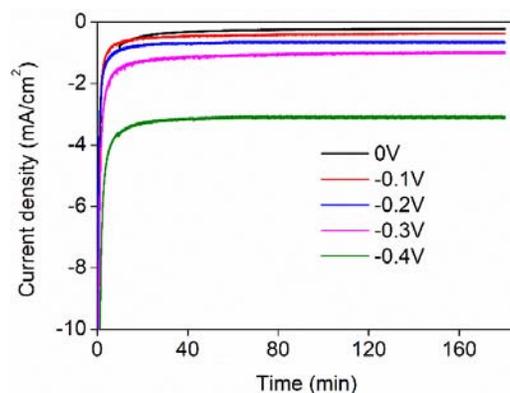

**Fig. S7. Temporal behavior of the electrolysis process of the NCM electrode at different potentials (vs RHE).**

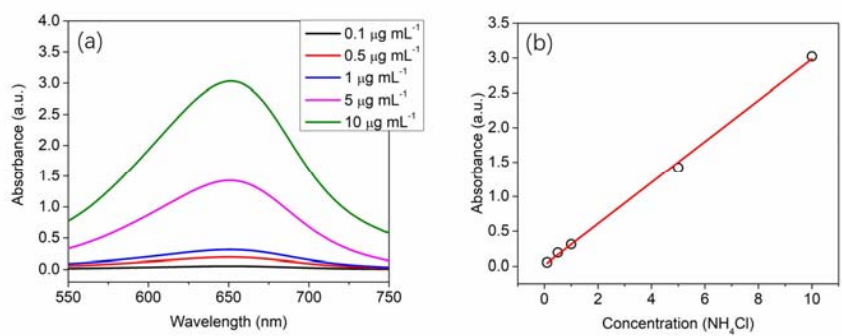

**Fig. S8. Absolute calibration of the indophenol blue method using ammonium chloride solutions of known concentration as standards.** (**A**) UV-Vis curves of indophenol assays with $NH_4^+$ ions after incubated for 2 hours at room temperature; (**B**) calibration curve used for estimation of $NH_3$ by $NH_4^+$ ion concentration. The absorbance at 655 nm was measured on a UV-Vis spectrophotometer, and the fitting curve shows good linear relation of absorbance with $NH_4^+$ ion concentration (**y = 0.2972x + 0.0134, R2=0.9986**) for three independent calibration curves.

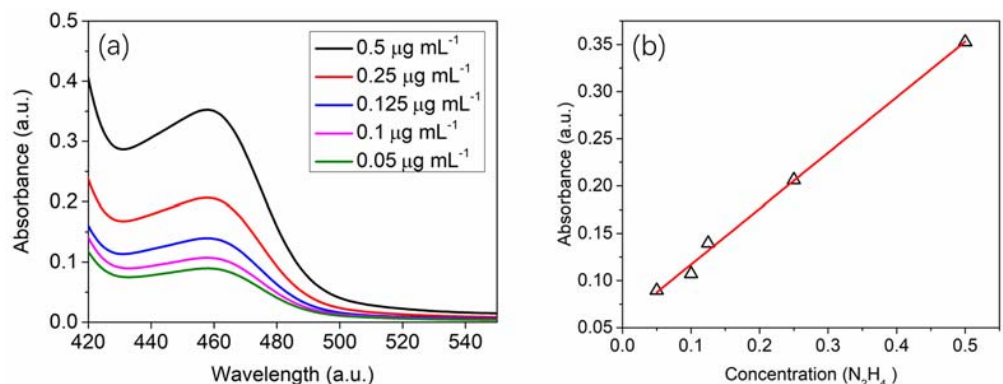

**Fig. S9. Absolute calibration of the Watt and Chrisp (para-dimethylamino-benzaldehyde) method for estimating $N_2H_4 \cdot H_2O$ concentration, using $N_2H_4 \cdot H_2O$ solutions of known concentration as standards.** (**A**) UV-Vis curves of various $N_2H_4 \cdot H_2O$ concentrations after incubation for 10 min at room temperature; (**B**) calibration curve used for estimation of the $N_2H_4 \cdot H_2O$ concentration. The absorbance at 455 nm was measured on a UV-Vis spectrophotometer, and the fitting curve shows good linear relation of absorbance with $N_2H_4 \cdot H_2O$ concentration (**y = 0.5898x + 0.0581, R2=0.9964**) for three independent calibration curves.

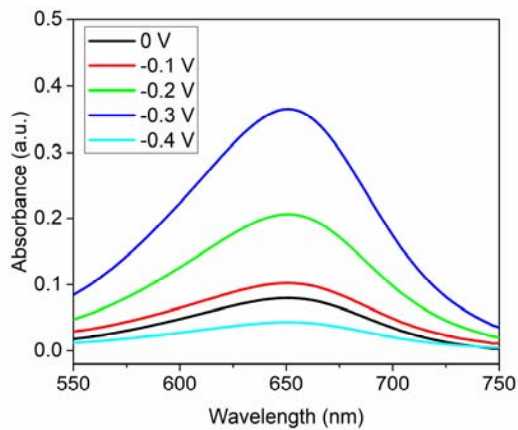

**Fig. S10.** UV-vis absorption spectra of the HCl electrolyte after electrolysis on the NCM electrode stained with indophenol indicator after charging at applied potentials vs. RHE for 3 h under various conditions.

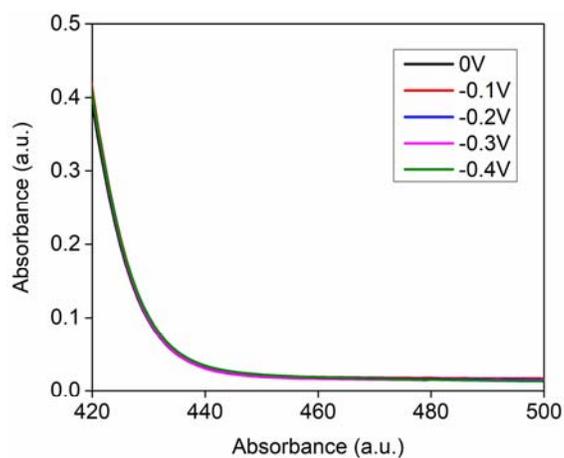

**Fig. S11.** UV-vis absorption spectra of the HCl electrolyte after electrolysis on a NCM electrode stained with indophenol indicator after charging at applied potentials vs. RHE for 3 h under various conditions.

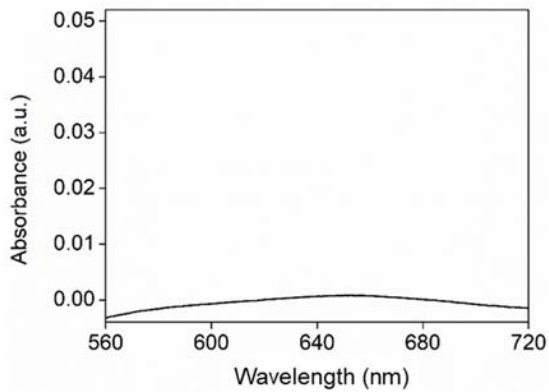

**Fig. S12.** UV-vis absorption spectra of the HCl electrolyte stained with indophenol indicator after charging on the NCM electrode at an applied potential of -0.2V (*vs*. RHE) under Argon bubbling for 3 h.

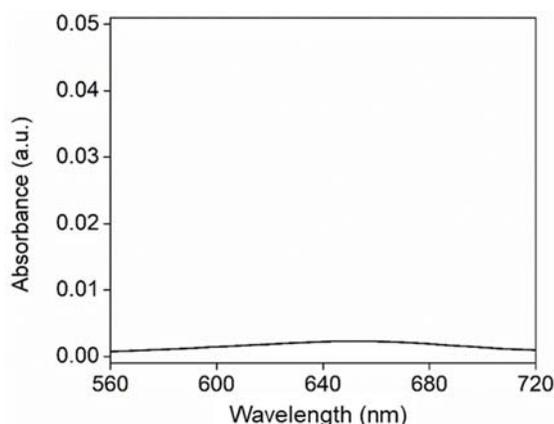

**Fig. S13.** UV-vis absorption spectra of the HCl electrolyte stained with indophenol indicator after charging on CNTs electrode at applied potential of -0.2 V (*vs*. RHE) under N$_2$ bubbling for 3 h.

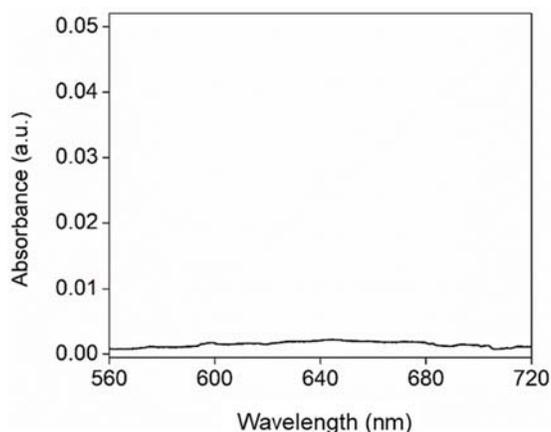

**Fig. S14.** UV-vis absorption spectra of the HCl electrolyte stained with indophenol indicator after charging on carbon paper electrode at applied potential of -0.2 V (*vs*. RHE) under N$_2$ bubbling for 3 h.

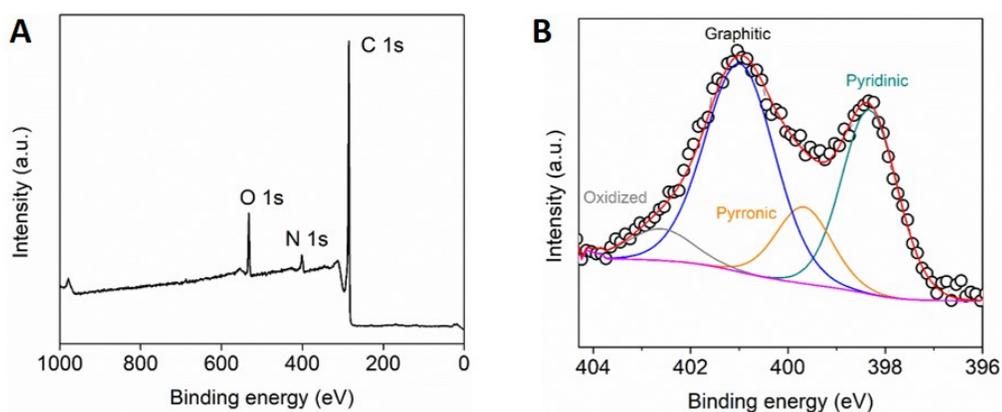

**Fig. S15. A,** XPS spectrum of the NCM after 4 days of electrolysis; **B,** High resolution N1 XPS spectra of the NCM after 4 days of electrolysis. Similar to the original HNDCM/CNT, the N species in HNCM/CNT after 4 days of electrolysis exist in the form of graphitic N (48.5%), pyridinic N (32%), pyrrolic (11.4%) and oxidized N (8.1%) produced, indicating its robust electrochemical stability.

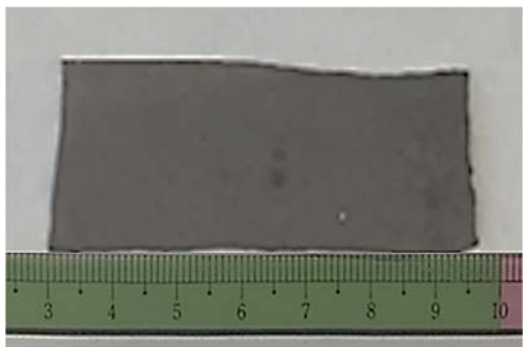

**Fig. S16. Photograph of a large piece of the NCM-Au NPs (unit: cm).**

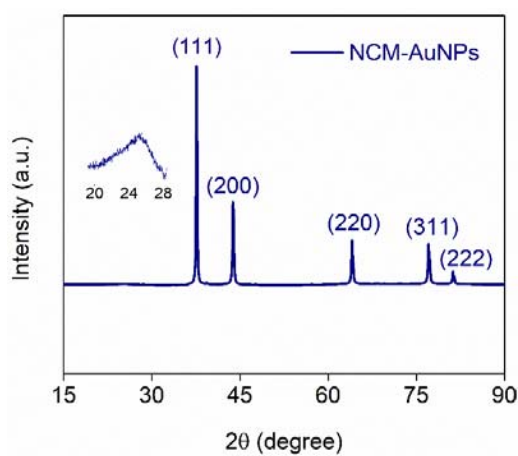

**Fig. S17. XRD pattern of the NCM-Au NPs. Inset is the (002) peak of graphitic carbon.**

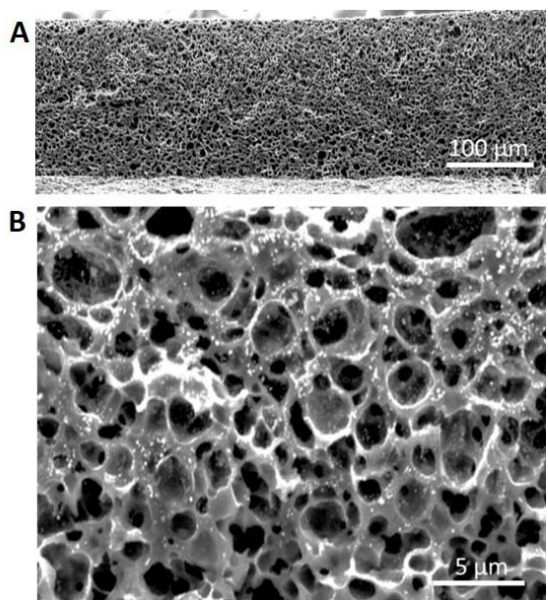

**Fig. S18. Low-magnification (A) and high-magnification (B) cross-section SEM images of NCM-Au NPs. From image B, it can be clearly seen that Au NPs are integrated within the NCM.**

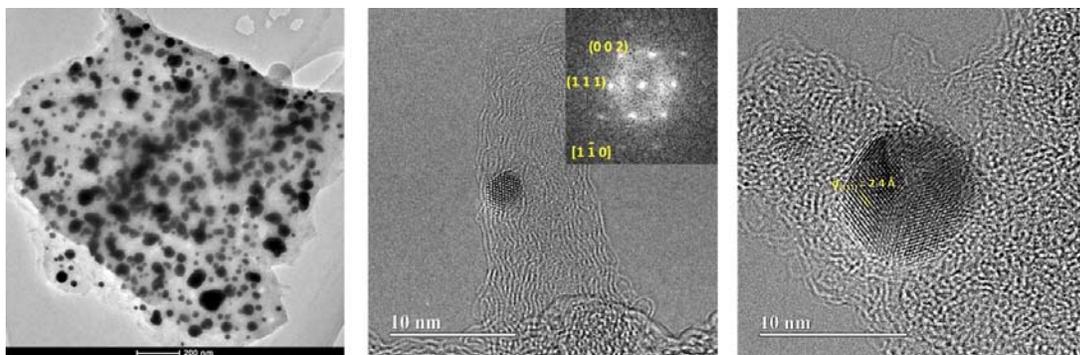

**Fig. S19. Low-magnification (a) and high-magnification (b) TEM images of the NCM-Au NPs. Inset in b is the selected-area electron diffraction (SAED) pattern for Au NPs embedded in NCM-Au NPs.**

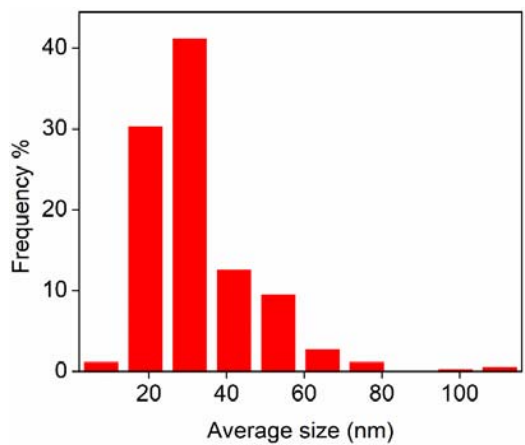

**Fig. S20. Histogram of the Au NPs size distribution on the NCM-AuNPs obtained from Fig S19a.**

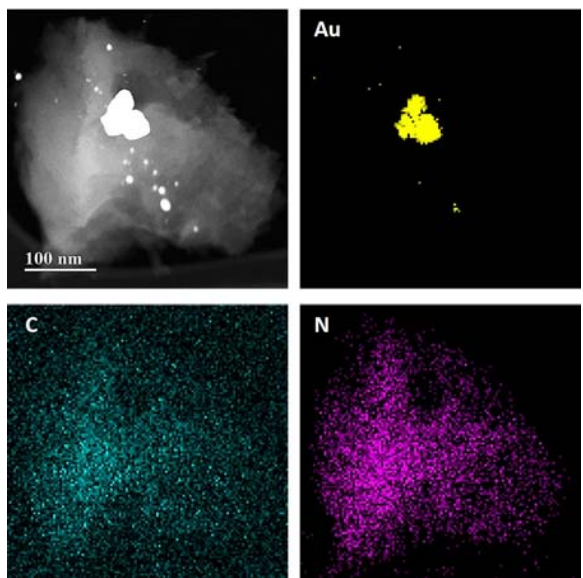

**Fig. S21. TEM image and corresponding elemental (Au, C and N) mappings.** A uniform distribution of N in the carbon matrix can be clearly seen, which is expected due to in situ molecular doping of N in NCM-AuNPs.

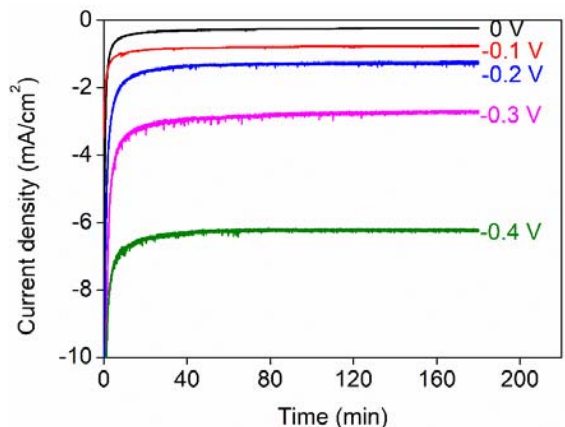

**Fig. S22. Temporal behavior of the electrolysis process of the NCM-Au NPs electrode at different potentials (vs RHE).**

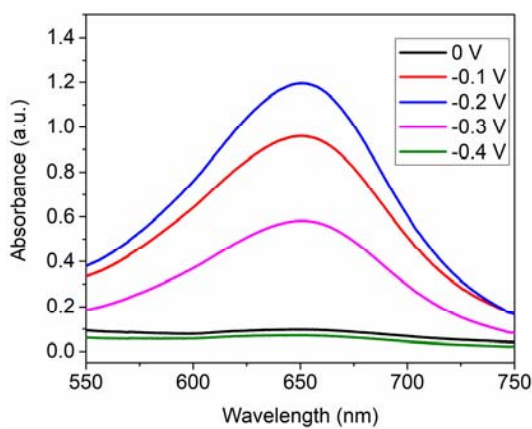

**Fig. S23. UV–vis absorption spectra of the HCl electrolyte after electrolysis on the NCM-Au NPs electrode stained with indophenol indicator after charging at applied potentials vs. RHE for 3 h under various conditions.**

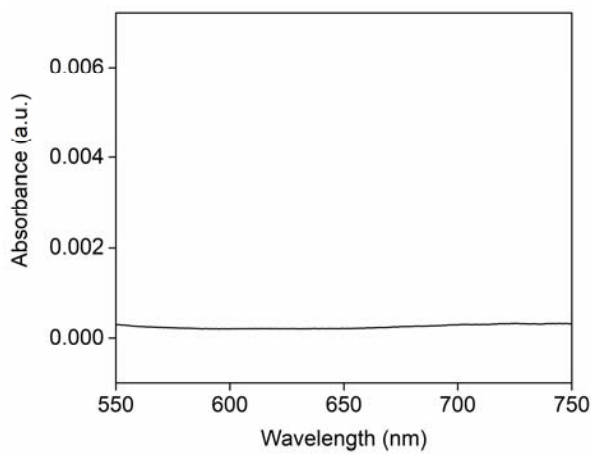

**Fig. S24. UV–vis absorption spectra of the HCl electrolyte after bubbling $N_2$ for 3h without employing applied potential.**

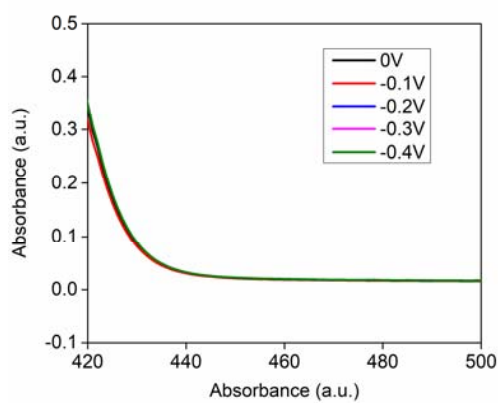

**Fig. S25. UV-vis absorption spectra of the HCl electrolyte after electrolysis on the NCM-Au NPs electrode stained with indophenol indicator after charging at applied potentials vs. RHE for 3 h under various conditions.**

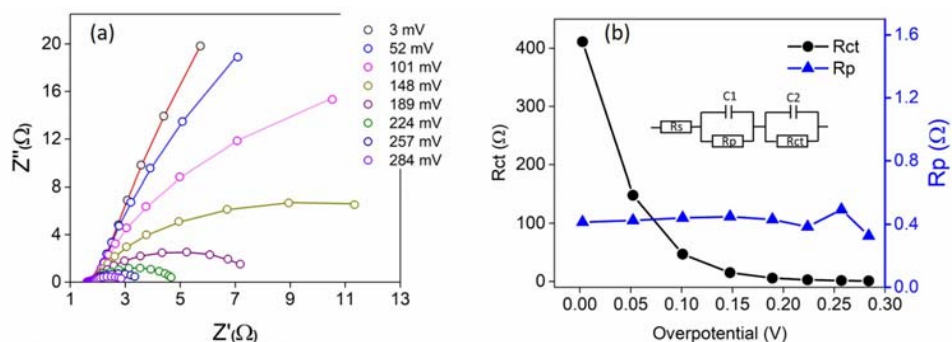

**Fig. S26.** (a) Nyquist plots of NCM-AuNPs for NRR at different overpotentials; (b) the fitting data of Rp and Rct are as a function of overpotentials in NRR, respectively. The semicircles in the high-to-low frequency ranges of the Nyquist plots could be attributed to the series resistance Rs, pore resistance Rp, and charge-transfer resistance Rct, respectively.

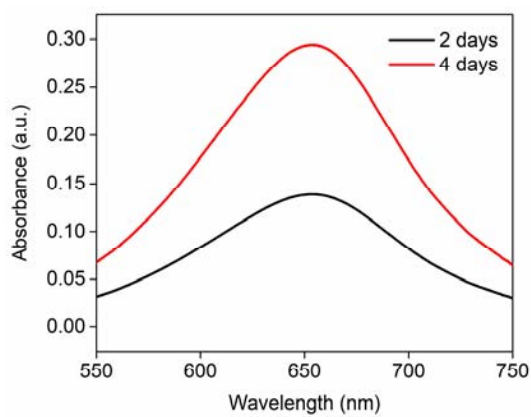

**Fig. S27. UV-vis absorption spectra of the HCl electrolyte after electrolysis on the NCM electrode stained with indophenol indicator after charging at applied potentials vs. RHE for different interval times at -0.1 V (vs RHE).**

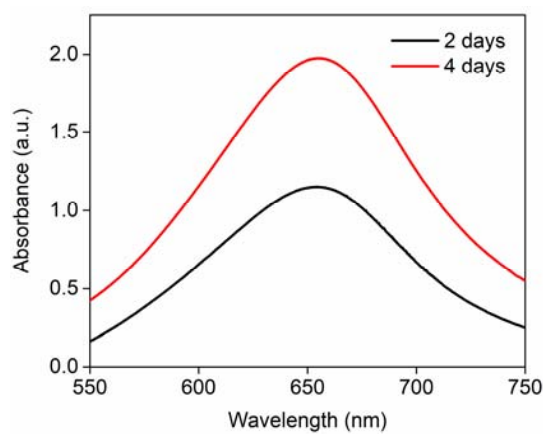

**Fig. S28.** UV-vis absorption spectra of the HCl electrolyte after electrolysis on the NCM-AuNPs electrode stained with indophenol indicator after charging at applied potentials vs. RHE for different interval times at -0.1 V (vs RHE).

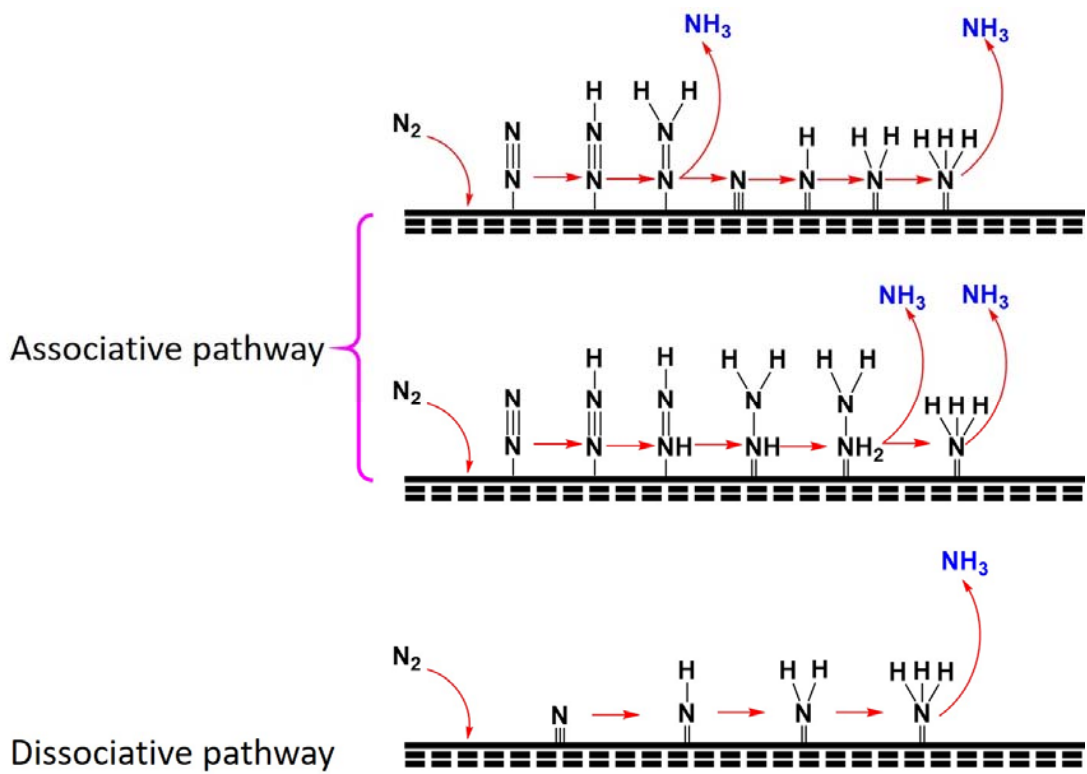

**Fig. S29.** Proposed mechanisms of NRR on the surface of a generic heterogeneous catalyst.

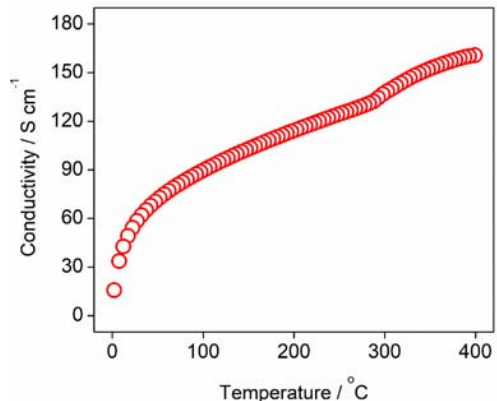

**Fig. S30. Temperature dependence of conductivity of NCM from 5 K to 390 K using a four-probe method.**

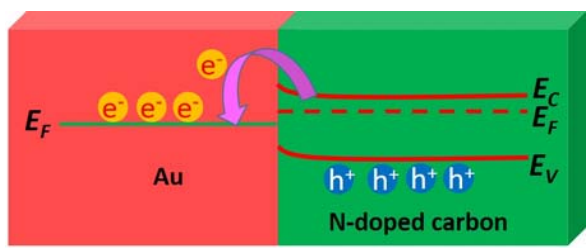

**Fig. S31. The rectifying effect arising from a metal/n-doped NCM junction.**

**Table S1**. Comparison of different electrocatalysts in NRR under ambient condition

| Catalyst | Overpotential (vs . RHE) | electrolyte | Faradaic Efficiency | Production yield (NH3 if is not specific) | reference |
|---|---|---|---|---|---|
| **N-doped Porous Carbon** | **-0.2 V** | **1M HCl** | **6%** | **(-0.3 V) 0.09 g m$^{-2}$ h$^{-1}$** | **Our work** |
| **Au nanoparticle @ N-doped Porous Carbon** | **-0.1 V** | **1M HCl** | **22%** | **(-0.2V) 0.36 g m$^{-2}$ h$^{-1}$** | **Our work** |
| MOF derived N-doped carbon | -0.3 V | 0.1 M KOH | 10.2% | 3.4 x10$^{-3}$ mmol cm$^{-1}$ h$^{-1}$ | Nano Energy, 2018, 48, 217-226 (S4) |
| MOF derived N-doped carbon | -0.9 V | 0.05 M H$_2$SO$_4$ | 1.40%[a] | 1.40 mmol g$^{-1}$ h$^{-1}$ | ACS Catal. 2018, 8, 1186−1191(S5) |
| Au nanoparticles @ CeOx-RGO | -0.2 V | 0.1 M HCl | 10.10% | 8.3 μg h$^{-1}$ mg$^{-1}$ cat. | Adv. Mater. 2017, 1700001 (S6) |
| Tetrahexahedral gold nanorods | -0.2 V | 0.1 M KOH | 4.02% | NH$_3$ : 1.648 μg h$^{-1}$ cm$^{-2}$ N$_2$H$_4$ H$_2$O: | Adv. Mater. 2017, 29, 1604799 (S7) |

| | | | | 0.102 µg h$^{-1}$ cm$^{-2}$ | |
|---|---|---|---|---|---|
| Au Sub-nanoclusters @ TiO$_2$ | -0.2 V | 0.1 M HCl | 8.11% | 21.4 µg h$^{-1}$ mg$^{-1}$ cat. | Adv. Mater. 2017, 29, 1606550 (S8) |
| Li$^+$ incorporated poly(N-ethyl-benzene-1,2,4,5-tetracarboxylic diimide) | -0.5 V | 0.5 M Li$_2$SO$_4$ | 2.58% | 1.58 µg h$^{-1}$ cm$^{-2}$ | J. Am. Chem. Soc. 139, 9771-9774 (S9) |
| VN nanowire array on carbon cloth | -0.3 V | 0.1M HCl | 3.58% | 2.48 x 10$^{-10}$ mol$^{-1}$ s$^{-1}$ cm$^{-2}$ | Chem. Commun., 2018, DOI: 10.1039/C8CC00459E (S10) |
| N-doped carbon spikes | -1.19 V | 0.25 M LiClO$_4$ | 11.56% | 97.18 mg hour$^{-1}$ cm$^{-2}$ | Sci. Adv. 2018;4: 1700336 (S11) |

**References**

S1, Yuan, J., Giordano, C. & Antonietti, M. *Chem. Mater*, **2010**, *22*, 5003–5012.

S2, Bolt, D. F. Colorimetric Determination of Nonmetals, Vol. 2 (Wiley, 1978).

S3, Watt, G. W. & Chrisp, J. D. *Anal. Chem. 1952*, *24*, 2006-2008.

S4, S. Mukherjee, D. A. Cullen, S. Karakalos, K. Liu, H. Zhang, S. Zhao, H. Xu, K. L. More, G. Wang, G. Wu, *Nano Energy*, **2018**, *48*, 217-226

S5, Y. Liu, Y. Su, X. Quan, X. Fan, S. Chen, H. Yu, H. Zhao, Y. Zhang, J. Zhang, *Acs Catal*. **2018**, *8*, 1186-1191.

S6, S. Li, D. Bao, M. Shi, B. Wulan, J. Yan, Q. Jiang, *Adv. Mater*. **2017**, 1700001.

S7, D. Bao, Q. Zhang, F. Meng, H. Zhong, M. Shi, Y. Zhang, J. Yan, Q. Jiang, X. Zhang, *Adv. Mater*. **2017**, *29*, 1604799.

S8, M. Shi, D. Bao, B. Wulan, Y. Li, Y. Zhang, J. Yan, Q. Jiang, *Adv. Mater*. **2017**, *29*, 1606550.

S9, G. Chen, X. Cao, S. Wu, X. Zeng, L. Ding, M. Zhu, H. Wang, *J. Am. Chem. Soc*. **2017**, *139*, 9771-9774.

S10, X. Zhang, R. Kong, H. Du, L. Xia, F. Qu, *Chem. Commun*. **2018**, DOI: 10.1039/C8CC00459E.

S11, Y. Song, D. Johnson, R. Peng, D. K. Hensley, P. V. Bonnesen, L. Liang, J. Huang, F. Yang, F. Zhang, R. Qiao, A. P. Baddorf, T. J. Tschaplinski, N. L. Engle, M. C. Hatzell, Z. Wu, D. A. Cullen, H. M. Meyer III, B. G. Sumpter, A. J. Rondinone, *Sci. Adv*. **2018**, *4*: e1700336.